\documentclass[aps,prd,nofootinbib,twocolumn]{revtex4}

\usepackage{natbib}
\usepackage{graphicx}
\usepackage{color}
\usepackage{textcase}
\usepackage{amsmath, amsthm, amsfonts}
\usepackage{enumerate}
\usepackage{soul}
\usepackage[normalem]{ulem}
\usepackage[english]{babel}
\usepackage[T1]{fontenc}
\usepackage{amsmath}
\usepackage{hyperref}
\usepackage{multirow}
\usepackage{verbatim}
\usepackage{subfigure}

\definecolor{amber}{rgb}{1.0, 0.75, 0.0}
\newcommand{\pcd}[1]{\textcolor{blue}{#1}}

\makeatletter
\let\jnl@style=\rm
\def\ref@jnl#1{{\jnl@style#1}}

\def\aj{\ref@jnl{AJ}}                   
\def\actaa{\ref@jnl{Acta Astron.}}      
\def\araa{\ref@jnl{ARA\&A}}             
\def\apj{\ref@jnl{ApJ}}                 
\def\apjl{\ref@jnl{ApJL}}                
\def\apjs{\ref@jnl{ApJS}}               
\def\ao{\ref@jnl{Appl.~Opt.}}           
\def\apss{\ref@jnl{Ap\&SS}}             
\def\aap{\ref@jnl{A\&A}}                
\def\aapr{\ref@jnl{A\&A~Rev.}}          
\def\aaps{\ref@jnl{A\&AS}}              
\def\azh{\ref@jnl{AZh}}                 
\def\baas{\ref@jnl{BAAS}}               
\def\bac{\ref@jnl{Bull. astr. Inst. Czechosl.}}
\def\caa{\ref@jnl{Chinese Astron. Astrophys.}}
\def\cjaa{\ref@jnl{Chinese J. Astron. Astrophys.}}
\def\icarus{\ref@jnl{Icarus}}           
\def\jcap{\ref@jnl{J. Cosmology Astropart. Phys.}}
\def\jrasc{\ref@jnl{JRASC}}             
\def\memras{\ref@jnl{MmRAS}}            
\def\mnras{\ref@jnl{MNRAS}}             
\def\na{\ref@jnl{New A}}                
\def\nar{\ref@jnl{New A Rev.}}          
\def\pra{\ref@jnl{Phys.~Rev.~A}}        
\def\prb{\ref@jnl{Phys.~Rev.~B}}        
\def\prc{\ref@jnl{Phys.~Rev.~C}}        
\def\prd{\ref@jnl{Phys.~Rev.~D}}        
\def\pre{\ref@jnl{Phys.~Rev.~E}}        
\def\prl{\ref@jnl{Phys.~Rev.~Lett.}}    
\def\pasa{\ref@jnl{PASA}}               
\def\pasp{\ref@jnl{PASP}}               
\def\pasj{\ref@jnl{PASJ}}               
\def\ptep{\ref@jnl{PTEP}}               
\def\rmxaa{\ref@jnl{Rev. Mexicana Astron. Astrofis.}}%
\def\qjras{\ref@jnl{QJRAS}}             
\def\skytel{\ref@jnl{S\&T}}             
\def\solphys{\ref@jnl{Sol.~Phys.}}      
\def\sovast{\ref@jnl{Soviet~Ast.}}      
\def\ssr{\ref@jnl{Space~Sci.~Rev.}}     
\def\zap{\ref@jnl{ZAp}}                 
\def\nat{\ref@jnl{Nature}}              
\def\iaucirc{\ref@jnl{IAU~Circ.}}       
\def\aplett{\ref@jnl{Astrophys.~Lett.}} 
\def\apspr{\ref@jnl{Astrophys.~Space~Phys.~Res.}}
\def\bain{\ref@jnl{Bull.~Astron.~Inst.~Netherlands}} 
\def\fcp{\ref@jnl{Fund.~Cosmic~Phys.}}  
\def\gca{\ref@jnl{Geochim.~Cosmochim.~Acta}}   
\def\grl{\ref@jnl{Geophys.~Res.~Lett.}} 
\def\jcp{\ref@jnl{J.~Chem.~Phys.}}      
\def\jgr{\ref@jnl{J.~Geophys.~Res.}}    
\def\jqsrt{\ref@jnl{J.~Quant.~Spec.~Radiat.~Transf.}}
\def\memsai{\ref@jnl{Mem.~Soc.~Astron.~Italiana}}
\def\nphysa{\ref@jnl{Nucl.~Phys.~A}}   
\def\physrep{\ref@jnl{Phys.~Rep.}}   
\def\physscr{\ref@jnl{Phys.~Scr}}   
\def\planss{\ref@jnl{Planet.~Space~Sci.}}   
\def\procspie{\ref@jnl{Proc.~SPIE}}   
\def\rmp{\ref@jnl{Rev.~Mod.~Phys.}}  
\def\plb{\ref@jnl{Phys.~Let.~B}}  
\def\cqg{\ref@jnl{Class.~Quantum~Grav.}} 
\def\lrr{\ref@jnl{Living~Rev.~Rel.}} 

\makeatother

\begin{document}

\title{A New Method to Observe  Gravitational Waves emitted by  Core Collapse Supernovae}

\author{P. Astone$^{1a}$, P. Cerd\'{a}-Dur\'{a}n$^2$ , I. Di Palma$^{1a,b}$, M. Drago$^3$, F. Muciaccia$^{1a,b}$, C. Palomba$^{1a}$, F. Ricci$^{1a,b}$}
\affiliation{$^{1a}$ INFN, Sezione di Roma, I-00185 Roma, Italy}
\affiliation{$^{1b}$ Universit\`a di Roma  {\it{La Sapienza}}, I-00185 Roma, Italy}
\affiliation{$^2$ Departamento de Astronom\'ia y Astrof\'isica, Universitat de Val\`encia, 46100 Burjassot (Valencia), Spain}
\affiliation{$^3$  Gran Sasso Science Institute, 67100 L'Aquila and INFN LNGS, 67100 Assergi (L'Aquila), Italy}



\begin{abstract}
While gravitational waves have been detected from mergers of binary black holes and binary neutron stars, signals from 
core collapse supernovae,  the most energetic explosions in the modern Universe, have not been detected yet.
Here we present a new method to analyse the data of the LIGO, Virgo and KAGRA network to enhance the  detection efficiency of this category of signals. The method takes advantage of a peculiarity of the gravitational wave signal emitted in the core collapse supernova and it  is based on a classification procedure of the time-frequency images of the network data performed by a convolutional neural network  trained to perform  the task to recognize the signal.
We validate the method using phenomenological  waveforms injected in Gaussian noise whose spectral  properties are those of the LIGO and Virgo advanced detectors and we conclude that this method can identify the signal better than the present algorithm devoted to select gravitational wave  transient signal.
\end{abstract}

\maketitle

\section{Introduction}

The direct observation of gravitational waves (GWs) by the advanced kilometer-scale GW detectors, which have been  operative in the period between 2015 and  2017, is a major milestone in physics and astrophysics \cite{Observation,final_O1,BBH-June,first-triple-event,NS-NS-GW-event,GRB-GW,multimessenger}.  So far, all the observed GW signals have been produced at the merger of compact binary
systems. All but one correspond to black hole binaries with total mass in the range of tens of solar masses. The observation of the binary neutron star merger  in 2017 \citep{NS-NS-GW-event} is a crucial milestone of the multi-messenger astronomy because of the combined detection of  GW and electromagnetic observations \citep{GRB-GW,multimessenger}.

\noindent
In the future we expect  that  the LIGO, Virgo and KAGRA interferometers will observe  other astrophysical phenomena.  The collapse of the core of massive stars ($\sim 10-100\;M_\odot$), in particular those producing core-collapse supernovae (CCSNe), was considered  a potential source of detectable GWs already at the epoch of  resonant bar detectors.  GW are emitted by aspherical mass-energy dynamics that include quadrupole or higher-order gravitational contributions. If  this asymmetric dynamics is  present in the pre-explosion stalled-shock phase of CCSNe, we should have the chance to observe the violent death of massive stars also via the gravitational channel. GW bursts from CCSNe encode information on the core dynamics  of a dying massive star and may enlighten  the  mechanism driving supernovae. 
 
 Early analytic and semi-analytic estimates of the GW signature of stellar collapse and CCSNe gave optimistic signal strengths ($ \sim 10^{-2} M_\odot c^2$), while modern multi-dimensional  simulations predict emission frequencies in the band of ground-based laser interferometers ($ 10~Hz ~-~10 ~ kHz$)  with total emitted GW energies  in the range $10^{-12}  -  10^{-8} M_\odot c^2$. These predictions suggest that even advanced interferometers will only be able to detect GWs from CCSNe at  distances lower than  $1 -100~ kpc$  with an optimistic rate event of the order of $1/25~ yr$. 
 In other models of more extreme scenarios, involving non-axisymmetric rotational instabilities, centrifugal fragmentation and accretion disk, the emitted GW signals may be sufficiently strong to be detectable to  distances of $(10 -15)~ Mpc$. At this distance  the Virgo cluster is included in the sphere centered on Earth and, as consequence, the potential detection rate of the advanced detectors increases  up to values higher than $ 1~ yr$. 
 
 A  credible CCSNe scenario is based on a collapse of the star's iron core (see e.g. \citep{Janka:2012, Burrows:2013}, for recent reviews), 
 which results in the formation of a proto-neutron star (PNS) and an expanding hydrodynamic shock wave. 
 The shock gets immediately stalled by presence of a continuous accreting flow. On a timescale of $\sim 0.2-1\;s$, a 
 yet-uncertain supernova mechanism  revives the shock that reaches the stellar surface and produces the spectacular electromagnetic emission of a type-II or type-Ib/c supernova. 
 If the shock fails to revive, a black hole is formed with no  or  very weak electromagnetic signature associated. In this work we refer generically as CCSNe to any core collapse
 event, regardless if the final outcome is a supernovae or the formation of a black hole.  

The supernova type classification is based on the explosion light curve and spectrum, which depend largely on the nature of the progenitor star. The time from core collapse to breakout of the shock through the stellar surface and first supernova light is minutes to days, depending on the radius of the progenitor and energy of the explosion.

Any core-collapse event generates a burst of neutrinos that releases most of the proto-neutron star's gravitational binding energy ($\sim 10^{53}\;erg$ ${ \sim 0.15~M_\odot c^2}$) on a time scale of the order of $10\;s$. This neutrino burst was detected from SN 1987A and confirmed the basic theory of CCSNe \citep{Hirata:1987}.\\

Multi-dimensional simulations of core-collapse supernovae are currently at the frontier of research in the field following the two main basic explosion paradigms:
the {\it neutrino-driven} mechanism, thought  to be active for slowly rotating progenitors and responsible for the most common SNe, and the {\it magneto-rotational mechanism}, active only
for fast rotating-progenitors and responsible for rare but highly energetic events, like hypernovae and long GRBs. 

Several groups worldwide are currently attacking this problem with two- and three-dimensional simulations using the world's most powerful supercomputers. 
Multiple challenges arise during the numerical modelling: i) accurate solution of the neutrino transport equations during the evolution; ii) incorporation of the complete interactions
of electron, muon and tau neutrinos and their anti-particles with matter; iii) use of high resolution to resolve numerically fine structure features in the convective and turbulent flow around the proto-neutron star;
this  is of special importance for the development of magneto-rotational instabilities in fast-rotating progenitors; iv) accurate (general relativistic) description of gravity; v) use of sophisticated equations of state 
to describe the behaviour of matter at high densities. The different groups studying the problem use different approaches to tackle each of these challenges and, to this point, no one has carried out a 
definitive three-dimensional simulation including all the physical ingredients and with sufficiently high resolution to give the world-wide community confidence in the results. 

Despite of the problem complexity, these calculations give acceptable remnant neutron-star masses and predicted already  few distinct signatures of GW signals in both the time and frequency domains. The core-bounce signal is the part of the waveform which is best understood~\cite{Dimmelmeier:2002b} and it can be directly related to the rotational properties of the core~\cite{Dimmelmeier:2008, Abdikamalov:2014, Richers:2017}. However, fast-rotating progenitors are not common and their bounce signal will be probably difficult to observe in typical  galactic events, due to its high frequency and low amplitude.

In addition, during the post-bounce evolution of the newly formed proto-neutron star (PNS), the convection determines the excitation of highly damped modes in the PNS by accreting material and instabilities (SASI), with a peculiar  GW emission. In this case  the GW waveforms last for about $200-500~ms$ until the supernova explodes (see e.g \cite{Mueller}) or, in the case of black hole formation, the typical duration is $1~s$ or above \cite{Cerda-Duran:2013}. The peculiarity is that the signal  frequencies raise monotonically with time due to the contraction of the PNS, whose mass steadily increases. Characteristic frequencies of the PNS can be as low as $\sim100~Hz$, specially those related to g-modes \citep{Murphy,Mueller,Cerda-Duran:2013,Kon2,Kuroda,Andre} and SASI \citep{Cerda-Duran:2013,Kuroda,Andre}, which make them a  target for ground-based interferometers with the highest sensitivity at 
those frequencies. 

These information can be used in the search of the GW signal embedded in the detector noise, with the perspective to increase the confidence detection of signal emitted in the deeper universe.
In the case of  the  search of GW binary systems the dominant analysis technique  is  the matched filter, based on models computed in the general relativity (GR) framework. Then, the posterior probability distribution for the signal parameters are estimated from the noisy detector data using  probabilistic Bayesian methods. 
These  techniques can be used in the case of a detailed  prediction of the waveform, a case different to the present one. For this reason the approach used in the past for CCSNe relies on  fully unbiased algorithms, which don't require any assumptions about the GW morphology.  In general, these algorithms assign a loudness measure to each event, whose significance is evaluated by computing  the rate at which the background noise produces events of equal or higher loudness (false alarm rate, FAR). 

Currently, in the CCSNe case we can take advantage of the signal peculiarity, in particular that  associated to monotonically raise of the frequency related to the g-mode excitation.  The aim of this paper is to present  
a search strategy of events in coincidence in the advanced detector network, characterised by a raising monotonic behaviour in the time-frequency  plane, similar to the one observed in numerical simulations. 

This strategy is based on machine learning techniques. These are  tools   applied even to big chunks of data in different contexts, analysed with minimal human supervision and able to resolve ambiguity and tolerate unpredictability. In this framework  pattern recognition, seen as practical outcome of the machine learning technique which  divides  data into classes,  is a data analysis approach widely used for recognising  regularities in images. 
This approach has been tested already on GW data in particular for  the real time detection and the parameter estimation of binary black hole mergers \cite{Daniel}, \cite{powell}, \cite{Cuoco}, \cite{Gabbard}. Here we present  a method helpful for  the search of  signals associated to the supernovae explosions. 
In the following sections, after the discussion of the scientific problem of the detection of the transient signal due to the supernovae explosion, we will describe the phenomenological waveforms generated to simulate the CCSNe GW signal, the architecture of the convolutional neural network and the whole method developed to recognise the signal. Finally, to validate the method, we present the results obtained by injecting waveforms in Gaussian noise with the spectral behaviour of the LIGO and Virgo Advanced detectors.

\section{Phenomenological waveforms for CCSNe}

The first step of a signal search based on machine learning technique is to provide a training data set, where the GW signals are present. It follows that we have to produce  GW templates representative of CCSNe covering the parameter space of possible core collapse  events. 
At present the outcome of  multidimensional numerical simulations is a  limited set of GW waveforms because of  the massive amount of computational 
resources needed to produce each of them. In addition the  progenitors models used in these simulations can be biased:  
most of them are developed  with the aim to compare the model prediction with the observations of the supernovae  SN1987A 
or they are focused to the  case of fast-rotating progenitors,  a small fraction of the total number of
observable events.  Furthermore, due to the numerical challenge of these simulations and the various approximations used, it is unclear how close the existing numerical templates are to the actual GW signal for a specific
type of progenitor. Therefore, the  existing numerical templates  seems to have   just a partial   coverage of the CCSNe parameter space.

For this reason, to validate our search method,  we use a  more flexible approach. We have developed a parametrised phenomenological waveform designed to match  the most common features  observed in the  numerical models of CCSNe and we devote the next section to  present the simulation  used to produce the template bank, which covers a wider parameter space.

\subsection{Reference numerical simulations}
\label{sec:numsim}

We base our phenomenological templates on the numerical simulations by \citep{Murphy,Mueller,Cerda-Duran:2013,Yak,Kuroda,Andre}. In all these works,
the authors present the gravitational waves signals extracted from core-collapse simulations, plot spectrograms of the signal, and interpret those spectrograms in terms of excitation of modes (g-modes and SASI modes) and convection. A more detailed analysis and interpretation of the waveforms in terms of eigenmodes of the proto-neutron star (PNS) has been carried out by \cite{Sot,Forne,Morozova:2018,Sotani,Torres-Forne:2018}.
These simulations were performed in two and three dimensions  (2D/3D), using either a modified Newtonian potential or general relativity (XCFC approximation) and a neutrino treatment with different degrees of sophistication (from a simple leakage to Ray-by-Ray+ transport). The progenitors used are non-rotating stars (except for \cite{Cerda-Duran:2013}) with solar metallicity (except \cite{Cerda-Duran:2013} and some models of \cite{Mueller}) and correspond to zero-age main-sequence masses in the range $8-40\;M_\odot$. This kind of progenitor is most likely to
form type II supernovae and in some cases a failed supernova (unnovae, in which a black hole is formed). 
We focus in this work exclusively in this kind of progenitors.
A galactic supernova (or an unnova) is very likely to have such a non-rotating progenitor, so the features presented in these work are the most relevant for a possible detection.  We note that the fraction
of CCSNe associated to the collapse of rapidly rotating core is probably below $1\%$ (see discussion in \cite{Woosley:2006}).

Waveforms from the collapse of non-rotating progenitors have the next features identified by several of authors: \begin{enumerate}
\item \emph{Bounce signal:} in practice almost non existing. Only fast rotating models give a strong signal at bounce \cite{Dimmelmeier:2008, Abdikamalov:2014, Richers:2017}.
\item \emph{Prompt convection:} Some models show prompt convection right after bounce, which lasts for $50-100~ms$ at about $100~Hz$ (see e.g. \cite{Marek:2009,Murphy}). The amplitude of this signal
is currently under debate and it may depend on fine details of the numerical simulations and on the equation of state.
\item \emph{Excitation of g-modes of the PNS:} basically all simulations in the literature show this feature. Its frequency starts around $100~Hz$ and grows in time as the mass of the PNS grows creating a characteristic raising arch in the spectrogram. It may start right after the bounce or with some delay (up to $\sim200~ms$). The signal last until the onset of the explosion or the formation of the black hole. This signal has been identified
as the lowest order $l=2$ g-mode ($^2g_1$) of the inner core of the PNS \cite{Torres-Forne:2018}.
\item \emph{SASI modes:} SASI modes are observed in models in which the SASI is active \cite{Cerda-Duran:2013,Kuroda,Andre}. It starts at $\sim100~Hz$, usually with some delay after bounce, and its frequency grows in time, albeit at a lower pace than g-modes. Its frequency growth is close to linear rather than an arch.
\item \emph{Memory:} The explosion and the anisotropic neutrino emission, leave a low frequency signal in the range $\sim1-10~Hz$ (e.g. \cite{Marek:2009,Murphy}), usually described as a memory effect.
\end{enumerate}

\subsection{Parametrised templates}

\begin{figure*}
\includegraphics[width=0.9\textwidth]{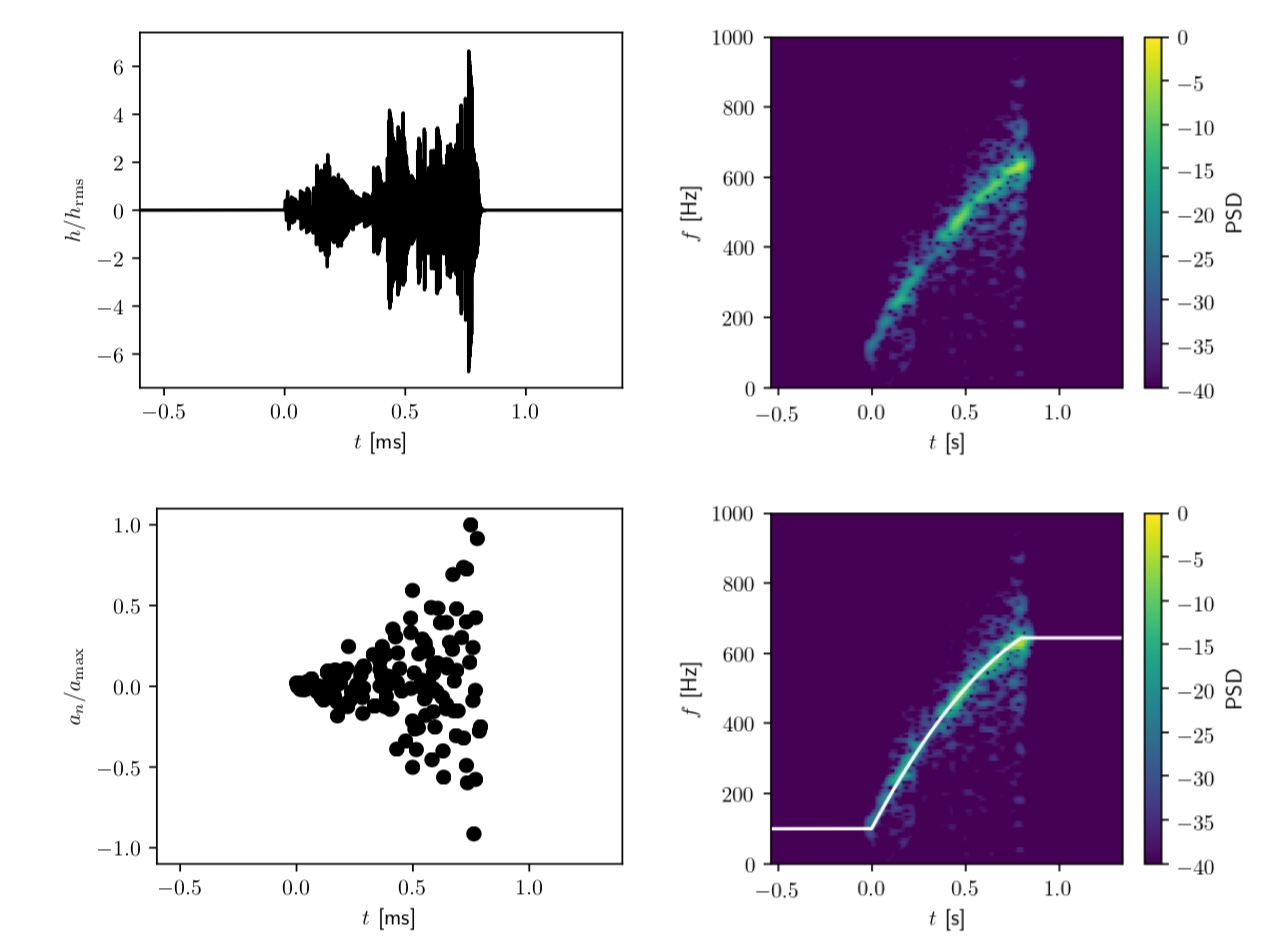}
\caption{Example of a realisation of a phenomenological template. We plot the strain (upper left),
the corresponding spectrogram (upper right), the coefficient of the impulsive acceleration (lower left) 
and the frequency of the harmonic oscillator over-plotted to the spectrogram (lower right). }
\label{fig:template}
\end{figure*}

We concentrate this work in the g-modes, the most common feature of all models, which also are responsible for the bulk of the GW signal in the post-bounce evolution of the PNS. 
The aim of our phenomenological template is to mimic the raising arch observed in core-collapse simulations. To this end we will consider a toy model for the GW emission in CCSNe. 
The idea is that at each time in the post-bounce evolution, the PNS is in quasi-hydrostatic equilibrium and any perturbation will excite the eigenmodes of the system, in particular g-modes. 
Non-spherical eigenmodes, in particular $l=2$ modes, will emit GWs at some characteristic frequencies corresponding to these eigenmodes.
This premise has been shown to be a quite accurate description of most of the waveforms \cite{Forne,Morozova:2018,Torres-Forne:2018}. These modes are continually being
excited by the hot bubble surrounding the PNS, in particular by convective motions and SASI \citep{Marek:2009,Murphy}. At the same time these excited modes are damped 
by the PNS conditions (e.g. by the existence of convective layers that do not allow for buoyantly supported waves) and by the presence of non-linearities and instabilities (e.g. turbulence).
Therefore, the GW emission can be modelled as a damped harmonic oscillator with a random forcing, in which the frequency varies with time. 

Following these arguments, we can model the strain measured at the detector as the solution of:
\begin{equation}
\partial_{tt} h + \frac{\omega(t)}{Q} \partial_t h + \omega(t)^2 h = a(t), \label{eq:harmonic}
\end{equation}
where $\omega(t)\equiv 2\pi f(t)$ is the angular frequency corresponding to the eigenmode excited in the PNS, $Q$ is the Q-factor, which we consider to be constant for simplicity,
and $a(t)$ is an acceleration driving the signal (the random forcing).

We model the frequency as a 2 degree polynomial:
\begin{equation}
f (t) = f_0 + f_1 (t-t_{\rm ini}) + f_2 (t-t_{\rm ini})^2 \qquad ; \quad t \in [t_{\rm ini}, t_{\rm end}],
\end{equation}
where $t$ refers to the post-bounce time and 
$f_0$, $f_1$ and $f_2$ are three coefficients determining the behaviour of the frequency evolution.
$t_{\rm ini}$ and $t_{\rm end}$ correspond to the beginning and end of the signal, being 
$t_{\rm end} - t_{\rm ini}$ its duration. Note that the beginning of the signal $t_{\rm ini}$ do not necessarily have
to coincide with the time of bounce ($t=0$), so it is possible to incorporate in the model the typical
delays (up to $200~ms$) observed in numerical simulations.

Instead of using $f_1$ and $f_2$ directly it is more convenient to define $f_{1s} \equiv f(t=1s)$ and $t_2$, the latter being the time at which the polynomial has a maximum. 
Given that the spectrograms of numerical simulations are not showing any maximum in the evolution of the features characterised as g-modes (at least in the pre-explosion phase),
the value of $t_2$ has to fulfil that $t_2 \ge t_{\rm end}$.
Using $f_0$, $f_{1s}$ and $t_2$ as parameters, the frequency can be expressed as
\begin{align}
f(t) &=& f_0 
+ \frac{2 (f_{1s} -f_0) (t_2 - t_{\rm ini})}{(2 t_2 - t_{\rm ini}- 1)(1-t_{\rm ini})}  (t-t_{\rm ini})\\
&&- \frac{f_{1s}-f_0}{(2 t_2 - t_{\rm ini}- 1)(1-t_{\rm ini})} (t-t_{\rm ini})^2,
\end{align}
where time is expressed in seconds.

To mimic the sudden downflows observed in numerical simulations responsible for the excitation of the g-modes, 
we model $a(t)$ as a series of $N$ instantaneous accelerations of the form $a_n \delta (t-t_{\rm n})$,
$n =1,...N$, with values of $t_n$ distributed randomly in the interval $[t_{\rm ini},t_{\rm end}]$ and with a random amplitude $a_n$ in the interval 
$[0,a_{\rm max}]$. $a_{\rm max}$ is  a normalisation constant, which we chose to be $\propto \omega^2$. There is an arbitrary constant
in the choice of $a_{\rm max}$, which is not of relevance for this work, because the templates are scalable to any desired amplitude. This 
constant could be calibrated in the future to generate distance-dependant templates, although this is beyond of the scope of this work. 
Also the dependence of this amplitude with $\omega$
should be explored in the future.

Finally, instead of using $N$ as a parameter for the template, we use:
\begin{equation}
f_{\rm driver} \equiv \frac{N}{t_{\rm end} - t_{\rm ini}},
\end{equation}
which is the driver frequency, i.e. the number of triggers per unit time
introduced by the forcing. Physically, $f_{\rm driver}$ is related to the characteristic frequency of the random perturbations exciting modes in the PNS. Since
these perturbations are expected to be driven by convection and SASI, its typical frequency is few hundred Hz.

In total we have the next set of $7$ free parameters for the parametrised phenomenological template: $t_{\rm ini}$, $t_{\rm end}$, $f_0$, $f_{1s}$, $t_2$,
$Q$ and $f_{\rm driver}$.  For a given set of parameters we solve numerically Eq.~(\ref{eq:harmonic}) by means of the first order symplectic-Euler method.
The computed waveform has some stochastic nature due to the randomness of the amplitude and time of the instantaneous accelerations. Therefore,
for a given set of parameters one can generate different realisations, depending on the seed used for the random number generator. This allow us to have
variability between waveforms corresponding to different realisations of the same model, something that has been observed when running  numerical 
simulations, e.g.  for simulations using different random perturbations in the initial model (see e.g. \citep{Summa:2006}).

All the waveforms generated for this work have a sampling rate of $20~kHz$, and have been padded with zeroes to a total duration of $2~s$
(in all cases longer than the duration of the signal) with the signal centered in the interval. To avoid errors from the numerical integrator we use a time-step
$10$ times smaller than the inverse of the sampling  rate.

An example of a waveform generated by our method can be seen in Fig.~\ref{fig:template}. The parameters used for this example are
$t_{\rm ini} = 0$, $t_{\rm end} = 0.8~s$, $f_0 = 100~Hz$, $f_{1s}=700~Hz$, $t_2=1.25~s$, $Q=10$ and $f_{\rm driver}=200~Hz$.
Note that the signal (upper left panel) only has power in regions where there are impulsive accelerations (lower left panel). Also the 
features in the spectrogram (upper right) follow closely the prescribed value for the frequency (lower right).
We note that the waveforms used in this study try to represent the typical features
observed in simulations of neutrino-driven CCSNe. This features are 
observed in numerical simulation by all groups in the community and their 
origin is well understood (g-modes in the proto-neutron star). This features
are not expected to disappear in future in more detailed numerical simulations, 
although the parameter space of possible values for the waveform 
may change in the future. A difference with respect to other works on the detection of CCSNe GWS
is that they make direct use 
of the waveforms from numerical simulations, while our approach allows to choose a relatively
wide parameter space that is able to encompass the results from future numerical simulations.
There is also room for improvement for the parametrised templates and we aim at making
a more comprehensive comparison between these templates and numerical waveforms, however
this is out of the scope of this work.

\subsection{Parameter space and template bank}

The range of possible values for the $7$ free parameters defining the waveforms
can be obtained by comparing with the spectrograms of the numerical simulations discussed in Section~\ref{sec:numsim}. The range of values
that we propose here are based on a simple inspection of the work by \citep{Murphy,Mueller,Cerda-Duran:2013,Yak,Kuroda,Andre}, 
with certain room such that we can accommodate any of these models inside our parameter space. The parameters $f_0$, $f_{2s}$
and $t_2$ are based in the inspection of the frequency evolution of the predominant feature in the spectrogram. $f_{\rm driver}$ is set 
to the typical values of SASI and convective motions frequency, as we argue above. The duration of the signal is based on the minimal and maximal 
duration of all waveforms from non-rotating progenitors (fast rotating progenitors can have a longer duration \cite{Pablo}). 
Finally, the parameter $Q$ controls the width of the feature in the spectrogram. Numerical simulations
show a wide variety of widths for these features. While some simulations show relatively narrow features (e.g. \cite{Mueller}), which would correspond to
$Q\sim10$, in other cases the signal in the spectrogram is very broad (e.g. \cite{Murphy}), corresponding to ($Q\sim1$). Note that $Q$ is limited
to values larger than $1/2$, otherwise the oscillations become overdamped. Table~\ref{tab:param} shows the parameter space explored in this work.

\begin{table}
\caption{Parameter space of the phenomenological templates. The second and third columns
indicate the range (maximum and minimum, respectively) for each parameter. The fourth 
shows the value used to generate the template bank in this work. Note
that not all combinations are possible since $t_{1s}>t_{\rm ini}$ has to be fulfilled.}
\begin{tabular}{l | c c | c }
\hline
parameter & min. & max.& test value\\ \hline
$t_{\rm ini}$ [s] & $0$ & $0.25$ & $0$ \\ 
$t_{\rm end}$ [s] & $0.2$ & $1.0$ & $0.5$\\ 
$f_0$ [Hz] & $100$ & $600$ & $100$\\ 
$f_{1s}$ [Hz] & $400$ & $2000$ & $1100$\\ 
$t_2$ [s] & $>t_{\rm end}$ & $\infty$ & $0.66$\\ 
$Q$  & $1$ & $10$ & $10$\\ 
$f_{\rm driver}$ [Hz] & $100$ & $600$ & $600$\\ \hline
\end{tabular}
\label{tab:param}
\end{table}

Given that this work is a proof-of-concept of the methods proposed, we use a single test value
within the parameter space (see Table~\ref{tab:param}), and we created
a template bank containing $100$ different realisations of this parameter set. 
This value is representative of a typical CCSNe waveform and in similar, e.g., to model M15 in  \cite{Mueller}. 
Therefore, the templates used in this work do not cover the all possible CCSNe scenarios and serve just as an example.
A deeper analysis covering the whole range of possible CCSNe scenarios will be developed elsewhere.


\section{The Method}
In this paper we use simulated data having the spectral behaviour of the advanced detectors LIGO and Virgo \cite{Curve}, with a standard Gaussian noise assumption.  We inject on these data the randomly generated waveforms described in the previous section. 
The SNR is defined as the square sum of the ratio of the reconstructed waveform in the frequency domain $(\tilde{h}_+,\tilde{h}_\times)$ and the amplitude spectral density $S_k(f)$ of each detector $k$:
\begin{equation}
SNR= \sqrt{\sum_k \int \frac{\tilde{h}_+^2+\tilde{h}_\times^2}{S_k(f)} df}
\end{equation}

Our method is essentially a two procedure steps:
\begin{itemize}
\item the data preprocessing derived in part from the software pipeline coherent Wave Burst (cWB) \cite{{Sergey}}, which prepares time-frequency images of the interferometer data;
\item a Convolutional Neural Network (CNN), which provides the classification of images in the noise or noise+signal classes;
\end{itemize}

\subsection{Data pre-processing}
The first step of the analysis is a pre-process based on the initial part of the pipeline coherent Wave Burst (cWB) of burst search.\\
cWB is the GW transient signal algorithm in use by the LIGO and Virgo collaborations that made the first alert of the GW150914 signal \cite{event}. It is an algorithm to measure energy excesses  over the detector noise in the time-frequency domain and combining these excesses coherently among the various detectors. This is performed introducing a maximum likelihood approach to define the ratio among the probability of having a signal in the data over the probability of only noise.
The algorithm is unbiased in the sense that does not depend on expected waveform, making it open to a wide class of transient signals.
The algorithm has been recently improved by implementing a new method of estimation of event parameters,  
which considers assumptions on the polarization state (circular, linear, elliptical, etc...) \cite{cWB2015}, \cite{DiPalma}.

\noindent
cWB looks for power excesses in the time-frequency domain using Wilsond-Daubechies-Meyer wavelet transform \cite{WDM}, which allows a better characterization of spectral features with respect to the Fourier transform.
The discrete wavelet transform are performed at different resolutions (wavelet levels), each one is an independent and complete representation of the original data. The likelihood approach allows to combine different time-frequency levels, having a unique wavelet representation adapted to the characteristic of the signals.\\
In our method, the pre-process is based on the wavelet transform applied to whitened data, since the time-frequency contains both the signal and the detector noise. 
cWB extracts from the network data a list of triggers above a defined threshold \cite{O1burst}.
\\
While in \cite{Vinciguerra} the time-frequency likelihood is used as input of a neural network for the identification of binary systems, here we apply a different approach more suited for CNN:  a fixed time-frequency level, i.e. the one which splits the available frequency band in 64 pixels, while the time size is fixed to 256 pixels.

The frequency upper limit is set to $1024~Hz$ as we did in \cite{O1burst}, and the 256 pixels correspond to a time window of $2~s$.
If the signal is too short in time to reach the number of 256, the pipeline includes adjacent pixels to image borders on left and right.
The extension on the left  is randomly  chosen in a uniform distribution between zero and the number of missing pixels, while the right one is the complement to the total number.

\begin{figure}[!h]
\begin{center}
\includegraphics[width=60mm]{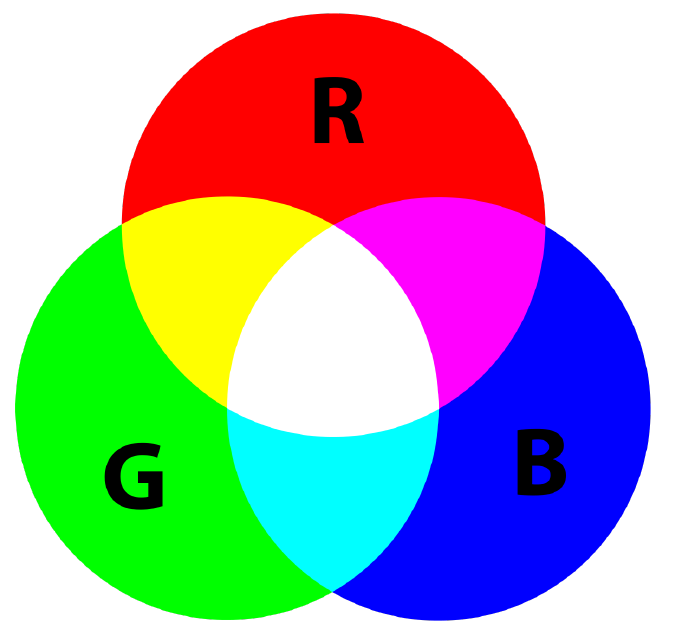}
\end{center}
\caption{\small{\textit{The mechanism of additive color synthesis. LIGO Hanford is assigned to red, LIGO Livingston to green and Virgo to blue. A triple coincidence will appear in white, while a double coincidence in yellow, magenta or cyan, depending on which couple of detectors is involved.}}}
\label{rgb}
\end{figure}

\begin{figure}[!h]
\begin{center}
\includegraphics[width=90mm]{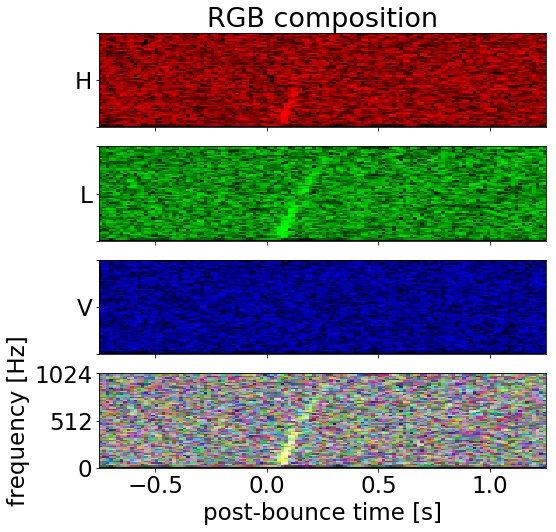}
\end{center}
\caption{\small{\textit{From the top; the spectrogram of LIGO Hanford is red, then that of LIGO Livingston is green and Virgo is blue. At the bottom: the RGB image obtained by stacking the previous three spectrograms. In this case, the signal is present just in Hanford and Livingston so that the combined signal at the bottom is in yellow.
}}}
%
\label{Fig. GridexampleL3}
\end{figure}

In practice, starting from the time domain data of the two LIGOs and one Virgo interferometer, we produce  three time-frequency sets of images, one for each detector.
Because the gravitational-wave signal must be present in at least two detectors we developed a technique to visually enhance the coincidences among all the interferometers of the network. The method consists in using  primary colours for the spectrograms of each detector: red (R) for LIGO-Hanford, green (G) for LIGO - Livingston and blue (B) for Virgo (see figure \ref{rgb}).
Then, the three single-colored spectrogram are stacked together to give as output an RGB image (see figure {\ref{Fig. GridexampleL3}}). 
The RGB spectrogram is a compact representation of the data to make evident the cross correlation between different detectors. This is an efficient way to prepare our data for the image recognition task that we will perform with our convolutional neural network.

\subsection{The Convolutional Neural Network}


Machine learning has become in recent years a cornerstone for many fields of science and it has been adopted more and more as a valuable asset \cite{Goo}. It has attracted much interest due to significant theoretical progress and due to the increased availability of large amounts of computing power (GPUs) and easy to use software implementations of standard machine learning techniques.\\

\noindent
CNN are designed to deal with spatially localised data, such as those found in images \cite{Fuk}. The network architectures of the CNN can be complex and include operations that go beyond those performed by the individual neurons of the networks. These extensions have allowed CNNs to become the state-of-the-art solution to several categories of problems, most notably photographic image classification \cite{EJ},\cite{Ver},\cite{Aur}.\\
Our aim is to provide a clear evidence that the machine learning technique, in particular neural network, can be more efficient to the respect of other approaches to extract GW CCSNe  signals, embedded in the detector noise and emitted in the far universe. 

The driving idea is to identify a  set of $N$ features in the data chunks, which are the outcomes of the CCSNe  3D simulations.
This set of information is  used to train a CNN that, thanks to its architecture, can proceed mostly in an automatic way  in the learning process. \\
In this architecture the neuron acts as an image filter and its weight can be thought as a specific pattern. For example, patterns might include different orientation of edges or small patches of colour. If the local region of the input matches that pattern, then the single neuron is activated. 
The input is scanned to look for the set of signal features and the process output is another image indicating where that pattern can be located.

\subsubsection{The CNN Architecture}
The model definition, the training and the validation phases have been developed  in the TFlearn \cite{TFL} and Keras frameworks \cite{Keras}, both based on the TensorFlow backend \cite{Flow}.

\begin{figure}[!h]
\begin{center}
\includegraphics[width=60mm]{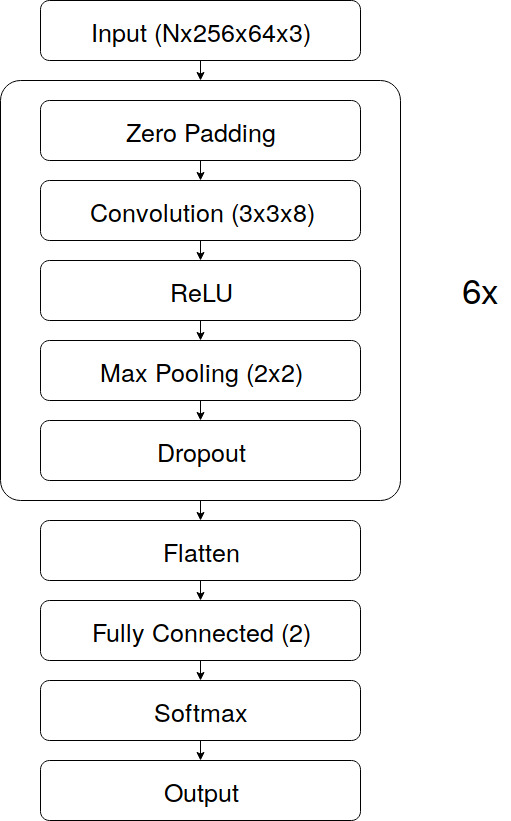}
\end{center}
\caption{\small{\textit{Sketch of the architecture of our model.}}}
\label{model}
\end{figure}

The network is designed as simple as possible while still having enough variables for the optimisation \cite{Haykin}. In figure \ref{model}, we sketch the block diagram of the CNN. The  images are inputs of the following sequence blocks: {\it{ ZeroPadding}},  {\it{ Convolutional}}, {\it{ Rectified Linear Unit ({\it{ReLU}})}}, {\it{ MaxPooling}}, {\it{ Dropout}}.

\noindent
The zero padding ensures that the size of the following convolutional output is still a power of two. Every convolutional layer in the network has the same
kernel size ($3\times 3$) and number of filters (8). 

\noindent
Every convolutional unit has a {\it {Rectified Linear Unit - ReLU}}, i.e. an activation function defined as:

$$ReLU(x)\;=\;max(0,x)$$

After the {\it{ReLu}} nonlinearity, a {\it{Max pooling}} \cite{Zhou} is performed. {\it{Max pooling}} is a downsampling process, which halves the image dimensionality,  it reduces the computational cost by decreasing  the number of parameters to learn and provides  local translation invariance to the internal representation. After the {\it{pooling}}, the minimum possible contraction to $2 \times 2$ implies that the feature map area is shrinked by a factor of four and  this operation returns the maximum output within a square neighbourhood. 

\noindent
The final step of every block is a soft dropout  aimed  to regularise the model and avoid over-fitting. 

\noindent
Then, the whole process is repeated six time, then flatten layer reshapes all the previous neurons in a one-dimensional vector.

This operation erases the information about topology, generally marking the
boundary between the convolutional and fully-connected part of the model. 
\noindent
After the flattening layer, we set a fully connected layer with only two output neurons, one for each class, \emph{noise} and \emph{signal+noise}. Those neurons have a {\it{softmax activation}} function, in order to obtain class probabilities as the final output of our classifier.  The softmax activation function is defined as:

\begin{equation}
softmax(\textbf{z})_i=\frac{e^{z_i}}{\sum_j e^{z_j}}
\end{equation}

where the indices $i$ and $j$ run from 0 to $n$ and the \textbf{z} array is the output of the $n$ neurons in the preceding layer.

\subsubsection{The learning phase and validation }
The output of the softmax layer is the probability $p$, while $q \equiv \{0,1\}$ is the true image class. 
\noindent
The learning phase is performed by a gradient descending algorithm of the loss function \cite{Goo} toward lower values.
To optimise our network we used the class of adaptive learning rate algorithms known as \emph{Adaptive-moments} (Adam) \cite{Kingma}, because of its robustness and fast convergence.

In a first phase the CNN  has to be trained to classify in the right class  the images containing the signal. Artificial neural networks have a huge number of internal parameters to adapt during the learning phase. In general, the more parameters involved the more expressive the network is. More expressive power means better ability to perform a given task, but also means longer and more difficult training, as well as higher computing resources needed.
With the architecture described before we have a total of 3210 partially correlated trainable parameters; this number is enough to represent the knowledge required to successfully perform our classification task.

For the training phase we split  the data   in two different chunks: the train and the test set. The model is trained only using the information
from the first set, while the second one is never involved in this process.  We feed the model with the train set while the test one is used to probe its generalisation capabilities, i.e.  it is effectively learning general features in our finite data set.
 The check of the learning process is done periodically to gain confidence on the classification efficiency performed on a completely new and independent set of data. Then, the information about this evaluation will be immediately discarded, without using them for the training: in this way, every successive evaluation will be like the first evaluation of a never-seen-before data set. The training phase stops when the gradient of the loss function is approaching zero within  our arbitrary chosen  interval of $10^{-4}$.\\


The {\it{ curriculum learning}}  starts the training from higher values  of the cost function ({\it{easier classification}}) and progressively decreases ({\it{harder classification}}). 
%
The training set is built using just images where we have signal-to-noise-ratio (SNR) higher than 4 in two detectors at least and we use different data set with decreasing SNR of the network: 40, 35, 30, 25, 20, 25, 20, 15, 12, 10, 8 
and, for each of this value, we compute the  CNN efficiency, $\eta_{CNN}$ , defined as 
\begin{equation}\label{Eq. effcnn}
\eta_{CNN} \;=\; \frac{\text{correctly classified signals }}{\text{all the signals at the input of CNN}}
\end{equation}
and the CNN False Alarm Rate ($FAR_{CNN}$) , defined as:
\begin{equation}\label{Eq. farcnn}
FAR_{CNN} \;=\; \frac{\text{misclassified noise }}{\text{all classified events}}
\end{equation}

\vskip 4mm

\begin{figure}[!h]
\begin{center}
\includegraphics[width=80mm]{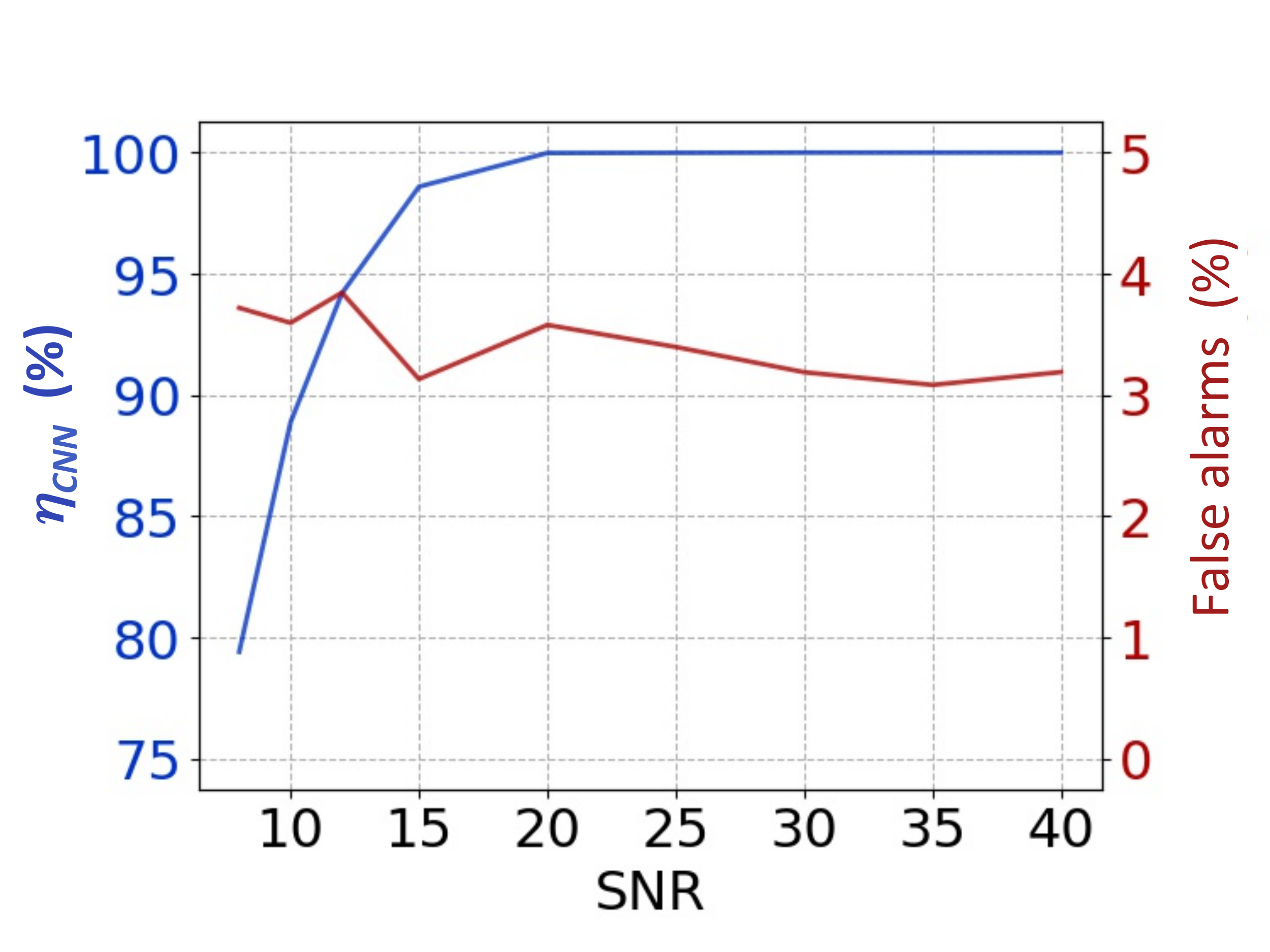}
\end{center}
\caption{\small{\textit{$\eta_{CNN}$ (Eq. \ref{Eq. effcnn}) for different SNRs (blue line) computed during the validation process at
threshold $\theta_{CNN}=0.5$. The red line is the false alarm probability ($\text{FAR}_{CNN}$ Eq. \ref{Eq. effcnn}) associated to the various SNRs computed at the same threshold. The used data set is based on 20000 signals and noise events for each SNR, half used for training and half for validation.
}}}
\label{Fig. Validation}
\end{figure}

The class encode $q$  of \emph{noise} and \emph{signal+noise} assumes just the value $0$ or $1$, while the output of the classifier is instead the real number $p$, the probability of the image to include a signal. Thus, to separate the two classes, we have to define the threshold $\theta_{CNN}$. The choice is the result of a trade off between the false positive and the false negative classification.
The main constraint is to minimise the signal dismissal, even if this implies to include some noise in the form of false positives.\\
\begin{figure}[!h]
\begin{center}
\includegraphics[width=90mm]{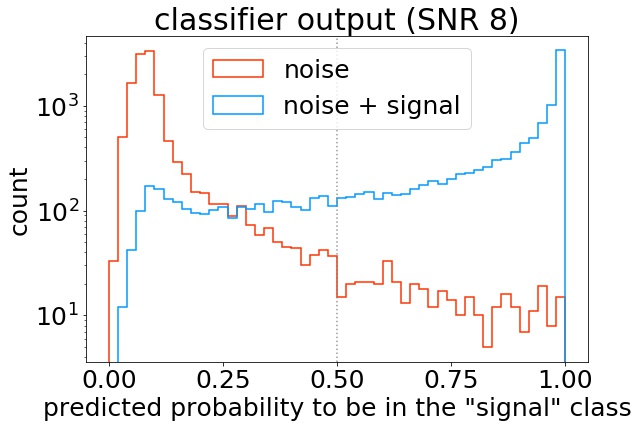}
\end{center}
\caption{\small{\textit{Distribution of the classifier output for noise  (red line) and  signal+ noise (light blue line) in the case of $SNR=8$}}}
\label{Fig. ClassOutput}
\end{figure}
In figure \ref{Fig. ClassOutput}, we show an example of class classification histograms of the noise and signal+noise  in the case $SNR=12$. 
By choosing a threshold of 0.5 in the predicted probability to be in the signal class (see fig \ref{Fig. ClassOutput}), we obtain the results shown in figure  \ref{Fig. Validation}, where for SNRs between 8 and 15  the efficiency is higher that $80 \%$, for SNR higher than $20$ is $1$ and the false alarm is confined in the range $3 \%$ and $4 \%$. 

\section{Results}
\begin{figure}[!h]
\begin{center}
\includegraphics[width=80mm]{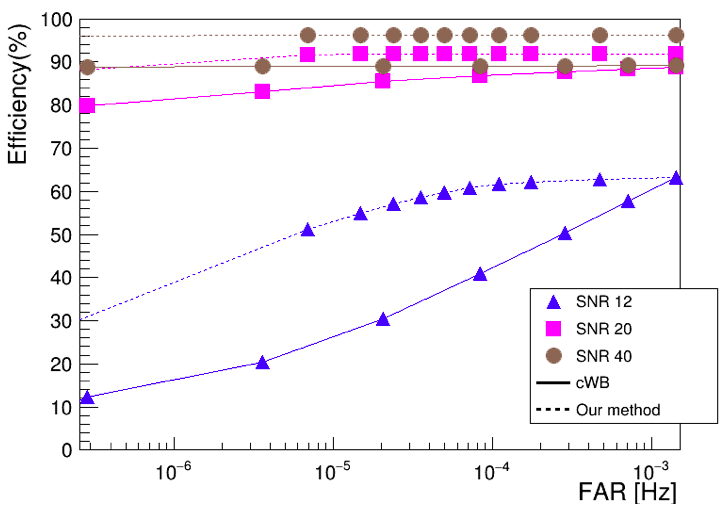}
\end{center}
\caption{\small{\textit{Efficiency vs false alarm for cWB (continuous line) and our method (dashed line) in the three cases SNR=12(blue triangles), SNR=20 (pink squares) and SNR=40 (brown circles).}}}
\label{Fig. ROC}
\end{figure}

In order to qualify the method, we compare its efficiency to the complete cWB procedure. The efficiency for
each procedure is defined as the ratio of the number of events passing the procedure thresholds and the number of injected events.
We simulate the background noise which is equivalent to six years of observation of advanced detectors, applying the usual time-shift procedure of the gravitational data analysis \cite{Sergey}. Then, we inject a subset of waveforms whose parameters are listed in Table~\ref{tab:param}. 
Since cWB's performances change in function of the SNR, in order to have compared number of events for each SNR value, we inject more signal at lower SNR.
\begin{table}
\begin{tabular}{|c|c|c|c|c|}
\hline
SNR & Injected & Input of CNN & Eff. cWB & Eff. our method\\
\hline
8 & 233994 & 16878 & 1.7 & 5.9\\ 
\hline
10 & 44946 & 14237 & 14.1 & 29.0\\ 
\hline
12 & 19500 & 12333 & 40.8 & 60.7\\ 
\hline
15 & 13500 & 11346 & 72.6 & 81.3\\ 
\hline
20 & 12000 & 11029 & 86.9 & 91.8\\ 
\hline
25 & 12000 & 11290 & 88.2 & 94.0\\ 
\hline
30 & 12000 & 11385 & 88.8 & 94.8\\ 
\hline
35 & 12000 & 11450 & 89.3 & 95.4\\ 
\hline
40 & 12000 & 11534 &  89.0 & 96.1\\ 
\hline
\end{tabular}
\caption{\small{\textit{Table of results at FAR=$7\cdot10^{-5}$ Hz. In the columns we report: SNR - the network SNR , Injected - the number of injections, Input of CNN - the number of injections found in the pre-processing stage by cWB and used as input of CNN, Efficiency - the efficiency of the different methods. }}}
\label{Tab. Validation}
\end{table}
For each SNR we build a set of 10000 time-frequency images and we combine them randomly with the same amount of noise images. The six years of observation time is  accordingly reduced to the image selection.\\
The comparison between our method and the complete cWB approach is done through the cross correlation statistics, $cc_{cWB}$, \cite{O1burst}. 
In particular, the post-production thresholds are set as reported in \cite{O1burst}, relaxing just $cc_{cWB}$ to  the value $0.6$, since we were not dealing with real noise.
We compute the efficiency curves of cWB and our method for every SNR as function of the false alarm rate. In figure \ref{Fig. ROC} we show just the case of SNR=12 and SNR=40 and we note that the efficiency of our method is better than that of complete cWB. Same results are obtained for all SNRs.

\begin{figure}[!h]
\begin{center}
\includegraphics[width=80mm]{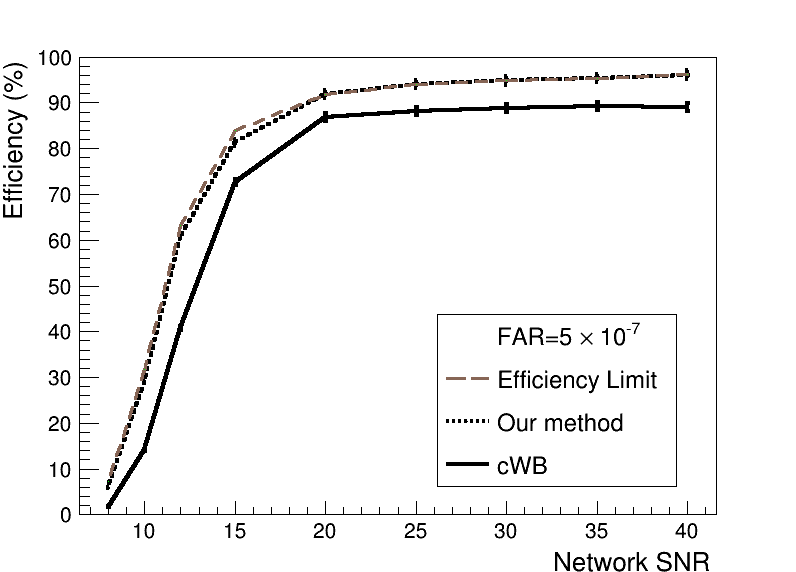}
\end{center}
\caption{\small{\textit{Efficiency vs SNR in the case of complete cWB (continuous) and our method (dashed) for all SNRs. We report also the curve that shows the ratio between the input events of the CNN and the total injected events in function of SNR (brown). This curve sets the maximum efficiency that our method can achieve}}}
\label{Fig. EffComp}
\end{figure}

In figure \ref{Fig. EffComp}, we plot the efficiency versus the signal to noise ratio of the network for the complete cWB and our method at the false alarm rate of about $7 \times 10^{-5}\; Hz$. 
Again, we note that our method has improved efficiency to respect to cWB. \\
In the same figure, we show also the ratio between the input events of the CNN and the total injected events in function of SNR, as listed in Table \ref{Tab. Validation}. This curve sets the maximum efficiency that our method can achieve. 
The missing events depend on the cWB post-production threshold, so that whole improvement is achievable by a better tuning of the cWB inputs.

\section{Conclusions}
We have presented a non-linear method based on convolutional neural network algorithm to extract CCSN signals embedded in Gaussian noise with spectral behaviour of Advanced LIGO and Virgo detectors. 
We compared the efficiency of the method for different signal to noise ratio to that of the algorithm used by the LIGO-Virgo collaborations to detect gravitational wave transient signals. The results show that our method has an higher efficiency and we conclude that using this new approach we can detect core collapse supernovae taking advantages of the peculiar features of the signal.\\
In the future, we plan to qualify the method using real detector data which are affected even by non-gaussian noise.
%

\section{Acknowledgments}
The authors gratefully acknowledge the supports of the Istituto Nazionale di Fisica Nucleare of Italy and of the {\it{Centro Amaldi}} of the department of Physics of the University of Rome {\it{Sapienza}} ( grant 000008-17 Ateneo2017). P. Cerd\'{a}-Dur\'{a}n acknowledges the support by the Spanish MINECO (grant AYA2015-66899-C2-1-P and RYC-2015-19074) and the Generalitat Valenciana (grant PROMETEOII-2014-069).

\end{document}